\documentclass[aps,
twocolumn,
showpacs,
amsmath,
amssymb,
nofootinbib]{revtex4} 

\begin{document}

\title{Holographic entanglement entropy of de Sitter braneworld}

\author{Yukinori Iwashita$^{(1)}$,  Tsutomu Kobayashi$^{(1)}$, Tetsuya Shiromizu$^{(1)}$
and Hirotaka Yoshino$^{(2)}$}


\affiliation{$^{(1)}$Department of Physics, Tokyo Institute of Technology, Tokyo 152-8551, Japan}
\affiliation{$^{(2)}$Graduate School of Science and Engineering, 
Waseda University, Tokyo 169-8555, Japan}

\date{\today}

\begin{abstract}
We study the holographic representation of the entanglement entropy,
recently proposed by Ryu and Takayanagi, in a braneworld context. 
The holographic entanglement entropy
of a de Sitter brane embedded in an anti-de Sitter (AdS) spacetime
is evaluated using geometric quantities, 
and it is compared with two kinds of de Sitter entropy:
a quarter of the area of the cosmological horizon on the brane 
and entropy calculated from the Euclidean path integral. 
We show that the three entropies coincide with each other in a certain limit. 
Remarkably, the entropy obtained from
the Euclidean path integral is in precise agreement with 
the holographic entanglement entropy in all dimensions.
We also comment on the case of a five-dimensional braneworld model
with the Gauss-Bonnet term in the bulk. 
\end{abstract}


\maketitle

\section{Introduction}

The exact description of the de Sitter entropy \cite{GH,GH2} is still 
an elusive problem \cite{dS,dS1}.
To clarify the microscopic origin of the de Sitter entropy is surely the first step toward understanding the fundamental nature of spacetime. 
Since de Sitter spacetime has a cosmological event horizon 
that has many formal similarities
with a black hole event horizon, it is 
natural to expect that quantum theory of black holes is 
closely related to that of de Sitter spacetime. 
Recent remarkable progress of string theory in many 
non-perturbative phenomena, such as strong/weak duality and the 
new solitonic objects, D-branes, elucidated the microscopic 
interpretation of black hole entropy \cite{SV}. 
However, there is severe adversity for discussing de Sitter gravity 
and the procedure of microstate counting breaks down in string theory. 
In this paper we address this issue in the braneworld context \cite{RS,dSbrane,dSbrane2}.

Recently Ryu and Takayanagi proposed the holographic derivation for 
the entanglement entropy \cite{RT}. 
The entanglement entropy emerges due to the existence of 
an inaccessible region. 
Consider $n$-dimensional static  
spacetime  whose $t={\rm const.}$ static slice 
is divided into two parts by 
a $(n-2)$-dimensional surface $S_{n-2}$. 
Through the anti-de Sitter/conformal field theory (AdS/CFT) 
correspondence \cite{adscft}, 
it is proposed that 
the entanglement entropy of ${\rm CFT}_n$ for accessible subspace is given by 
%
\begin{eqnarray}
S_{\rm ent}=\frac{A_{n-1}}{4G_{n+1}},
\end{eqnarray}
%
where $G_{n+1}$ is the gravitational constant in $(n+1)$-dimensional spacetime 
and $A_{n-1}$ is the area of a $(n-1)$-dimensional minimal surface 
$\Sigma_{n-1}$. The minimal surface $\Sigma_{n-1}$ 
is a submanifold in ${\rm AdS}_{n+1}$ whose boundary is given by $S_{n-2}$. 
This conjecture means that the entanglement entropy of CFT is given by the 
a holographic screen in dual AdS 
in the gravity side, which is called the holographic entanglement entropy. 
Ryu and Takayanagi confirmed the validity of the proposal for 
two-dimensional CFT and thermal CFT when applied to ${\text{AdS}}_3$. They 
also discussed the validity for higher dimensional cases. 
The proposal
is limited to static spacetime 
since their proposal is motivated by the holographic arguments in black 
holes~\cite{motivation} and 
the minimal surface is the horizon in static spacetime, 
although generalization to the stationary spacetime could be possible.

In the braneworld context, 
the proposal of the holographic entanglement entropy gives us the motivation for 
evaluating the entanglement entropy on the brane. 
Emparan applied their argument to the black hole localized on the brane, 
and confirmed that the holographic entanglement entropy is identical to the 
Bekenstein-Hawking entropy on the brane in a certain limit~\cite{Emparan}. 
It is greatly attractive to examine the holographic entanglement entropy 
for the braneworld black hole  because the minimal surface in the bulk 
is identical to the black hole event horizon. 
It also illuminates the mimic AdS/CFT 
correspondence in the Randall-Sundrum braneworld \cite{adscftbrane, tama}, 
where the bulk gravity dual is the
brane gravity coupled with CFT, rather than 
pure CFT. 
See Ref. \cite{afterword} for other applications.

In this paper, we address the relation between the holographic entanglement 
entropy and the other definitions of entropy of the de Sitter brane. We consider 
the $Z_2$ symmetric braneworld model. The problem is 
reduced to a purely geometrical problem, that 
is, finding the bulk minimal surface with the fixed boundary 
which is the cosmological horizon on the de Sitter brane. 
Although the minimal surface with a fixed boundary is not 
orthogonal to the brane in general, 
we can easily show the orthogonality thanks to the $Z_2$ symmetry across the brane. 
See Ref. \cite{HMS} for a similar argument in the lower dimensional 
braneworld. We also comment on the application to the braneworld 
model with the Gauss-Bonnet term, which is 
regarded as a string correction to low energy effective theory. 

The rest of this paper is organized as follows. In the next section, we 
compute the holographic entanglement entropy in the ($n+1$)-dimensional 
Randall-Sundrum braneworld with the $n$-dimensional de Sitter brane, 
and show that it agrees with the de Sitter entropy on the brane 
derived by the Euclidean path integral. 
Sec. III gives the discussion for the higher derivative theory in five dimensions.
In Sec. IV, we summarize our results and discuss the future problems. 
In Appendix A, we describe the derivation of the minimal surface 
in the bulk that is needed 
for the calculation of the holographic entanglement entropy.

\section{Holographic entanglement entropy of 
de Sitter braneworld}

The system we consider here is the $n$-dimensional de Sitter brane 
in $(n+1)$-dimensional AdS spacetime. The ($n+1$)-dimensional bulk 
metric is 
%
\begin{eqnarray}
 ds^2  
	& = & dr^2+(l H)^2{\rm sinh}^2(r/l)\times \nonumber \\
     & & \left[-dt^2+H^{-2}{\rm cosh}^2(Ht)d \Omega_{n-1}^2\right] \nonumber \\
	&=&  dr^2+(lH)^2{\rm sinh}^2(r/l)\times \nonumber \\
     && \left[ -(1-H^2\rho^2)dT^2 +\frac{d \rho^2}{(1-H^2\rho^2)}
	+\rho^2 d \Omega_{n-2}^2 \right],
\end{eqnarray}
%
where $l$ is the bulk AdS curvature length and
$H$ is the Hubble parameter on the brane located at $r=r_0$. 
The relation between static coordinates and global coordinates in dS is found in \cite{dS0}. 
The cosmological event horizon is located at $\rho=1/H$. 
$H$ can be written as
%
\begin{eqnarray}
H^{-1}=l ~{\rm sinh}(r_0/l). \label{hubble}
\end{eqnarray}
%
The ($n+1$)-dimensional 
gravitational constant is related to the $n$-dimensional one as \cite{dSbrane2}
%
\begin{eqnarray}
G_n=\frac{n-2}{2l}{\rm coth}(r_0/l) G_{n+1}. 
\end{eqnarray}
%
The brane tension $\sigma$ is related to the AdS curvature scale through 
Israel's junction condition as 
%
\begin{eqnarray}
\sigma = \frac{(n-1)}{4\pi G_{n+1} l}{\rm coth}(r_0/l).
\end{eqnarray}
%

Following Ref. \cite{RT}, the entanglement entropy is given by 
%
\begin{eqnarray}
S_{\rm ent}:=\frac{A_{n-1}}{4G_{n+1}},
\end{eqnarray}
%
where $A_{n-1}$ is the area of the bulk minimal surface on $T=$ constant 
static slices. As shown in Appendix A, the bulk minimal surface is the 
$(n-1)$-dimensional 
surface with $\rho=H^{-1}$. Its area is given by
%
\begin{eqnarray}
A_{n-1}=2\Omega_{n-2}l^{n-2} \int^{r_0}_0 dr~ {\rm sinh}^{n-2}(r/l),
\end{eqnarray}
%
where $\Omega_{n-2}$ is the area of a $(n-2)$-dimensional unit sphere, $\Omega_{n-2}=2\pi^{\frac{n-1}{2}}
/\Gamma(\frac{n-1}{2})$. Then the holographic entanglement entropy is evaluated as 
%
\begin{eqnarray}
S_{\rm ent}= \frac{(n-2)\Omega_{n-2}l^{n-2}}{4G_n}{\rm coth}(r_0/l)
\int^{r_0/l}_0dx~ {\rm sinh}^{n-2}x. \label{ent}
\end{eqnarray}
%

The de Sitter entropy for $n$-dimensional de Sitter 
spacetime is given by \cite{GH,GH2}
%
\begin{eqnarray}
S_{{\rm area}}=\frac{A_{\rm ch}}{4G_n}=\frac{\Omega_{n-2}}{4G_n H^{n-2}}. \label{dse}
\end{eqnarray}
%
In general $S_{\rm ent}$ is not equal to $S_{\rm area}$. For $r_0/l \gg 1$, 
however, the relation between the AdS curvature length and the 
cosmological horizon radius reduces to $e^{r_0/l}\simeq 2(Hl)^{-1}$
recalling Eq. (\ref{hubble}), and so we have 
%
\begin{eqnarray}
S_{\rm ent} \simeq \frac{\Omega_{n-2} l^{n-2}}{4G_n}\left(\frac{e^{r_0/l}}{2}\right)^{n-2} \simeq 
\frac{\Omega_{n-2}}{4G_n H^{n-2}}=S_{{\rm area}}. 
\end{eqnarray}
%
Thus the holographic entanglement entropy coincides with 
the de Sitter entropy on the brane in the limit of $r_0/l\gg 1 $. 
The discrepancy for general $r_0$ is not surprising 
because the gravitational field equations on the 
brane are different from the standard Einstein equations due to the brane/bulk interaction~\cite{SMS}, 
and hence the correction term to the entropy should be taken into 
account. 

The simple way to evaluate the de Sitter entropy on the brane 
taking such effect into account is to use 
the Euclidean path integral \cite{GH2}. If the gravity theory is 
weakly coupled, the dominant contribution to the path integral is made 
by those which are an extremum of the action and, then, the partition 
function can be adequately approximated by the classical action, 
$Z=e^{I_{\rm E}}$, where $I_{\rm E}$ is the Euclidean action. 
In the braneworld, we stress that $n$-dimensional computation corresponds  
to $(n+1)$-dimensional Euclidean path integral. Because the 
n dimensional effective action on the brane is given by the 
integration of the $(n+1)$-dimensional action over the coordinate 
of the extra dimension. Therefore, we should compute the 
$n$-dimensional de Sitter entropy by $(n+1)$-dimensional action. 
Accordingly, the entropy is given by 
%
\begin{equation}
S_{{\rm E}}=\beta \langle E \rangle -I_{\rm E}= -I_{\rm E},
\end{equation}
%
where the total energy 
$\langle E \rangle$ vanishes in the present case. 
Note that we should take the total action for the partition function because the bulk gravity 
contributes to the effective theory on the brane as mentioned above. 
However, because the bulk is static and has no event horizon, the entropy obtained by this partition function surely 
corresponds to the entropy which stems from the cosmological horizon on the brane. 

The total action $I_{\rm E}$ is evaluated as 
%
\begin{eqnarray}
I_{\rm E} & = & \frac{1}{16\pi G_{n+1}}\int d^{n+1} x {\sqrt {g}}\left[-\frac{n(n-1)}{l^2}-R \right] 
\nonumber \\
& & +  \int d^nx {\sqrt {q}}\left[ \sigma - \frac{1}{8\pi G_{n+1}}K\right] \nonumber \\
& = & \frac{n}{8\pi G_{n+1}l^2} \int d^{n+1}x {\sqrt {g}}
-\frac{1}{n-1}\sigma \int d^n x {\sqrt {q}} \nonumber \\
& = & \frac{n-2}{4(n-1)}\frac{l^{n-2}}{G_n}{\rm coth}(r_0/l)
\Biggl[n \int^{r_0/l}_0dx~{\rm sinh}^n x \nonumber \\
& & -{\rm coth}(r_0/l){\rm sinh}^n(r_0/l)\Biggr]
\Omega_{n-2}\nonumber \\
& = & - \frac{(n-2)\Omega_{n-2} l^{n-2}}{4G_n}{\rm coth}(r_0/l)
\int^{r_0/l}_0dx ~{\rm sinh}^{n-2}x \nonumber \\
& = & -S_{\rm ent}, 
\end{eqnarray}
%
where $g$ and $q$ are the determinant of 
Euclidean metric of the bulk and the brane, respectively, and
$R$ is the Ricci scalar of the bulk spacetime.
We can see that $S_{{\rm E}}$ 
agrees exactly with the holographic entanglement entropy. 

It is expected that the difference between 
$S_{{\rm area}}$ and $S_{\rm ent}=S_{{\rm E}}$ 
comes from the higher order corrections, i.e., the Kaluza-Klein 
corrections to the brane theory. To see this, we focus on the $n=4$ case. 
From Eqs.~(\ref{ent}) and~(\ref{dse}) the difference
between $S_{\rm ent}$ and $S_{{\rm area}}$ is evaluated as 
%
\begin{eqnarray}
\frac{S_{\rm ent}-S_{{\rm area}}}{S_{{\rm area}}} \simeq (Hl)^2 {\rm log}(Hl/2)+\cdots. \label{diff}
\end{eqnarray}
%
The effective action $I_{\rm eff}$ on the brane is given by~\cite{adscftbrane}
%
\begin{eqnarray}
I_{\rm eff} & \simeq & 
\frac{1}{16\pi G_4}\int d^4 x{\sqrt {-q}}
\Bigl[{}^{(4)}R+ 4 l^2 {\rm log}(Hl/2){\cal R}_2  \Bigr] \nonumber \\
& & -\int d^4x {\sqrt {-q}}(\sigma-\sigma_{\rm RS})+\Gamma_{\rm CFT},
\end{eqnarray}
%
where ${\cal R}_2 := -(1/8){}^{(4)}R^\mu_\nu {}^{(4)}R^\nu_\mu+(1/24){[{}^{(4)}R]}^2$
with the Ricci tensor in the bulk ${}^{(4)}R^\mu_\nu $,
$\sigma_{\rm RS}$ is the Randall-Sundrum canonical tension defined by 
$\sigma_{\rm RS}:=3/(4\pi G_5 l)$ \cite{RS}, and $\Gamma_{\rm CFT}$ is the 
effective action for CFT on the brane. From the trace part of 
the effective field equations, we see 
%
\begin{eqnarray}
{}^{(4)}R & = & -8\pi G_4 \left[T_{\rm CFT}-4(\sigma-\sigma_{\rm RS}) \right] 
\nonumber \\
& =  & 2\left[-\frac{6}{l^2}+\frac{1}{6}(8\pi G_5 \sigma)^2  \right] \nonumber \\
& = & 12H^2,
\end{eqnarray}
%
where we used the fact that the trace of the effective stress tensor of CFT, $T_{\rm CFT}$, 
is given by the 
trace anomaly and the first line of the above equation is equivalent to the 
trace part of the 
effective equations derived by the geometrical projection \cite{SMS}.  
Therefore the Euclidean effective action gives us the 
entropy 
%
\begin{eqnarray}
S_{\rm ent} & \simeq & S_{{\rm area}}\left[ 1+\frac{2}{3}l^2H^{-2}{\rm log}(Hl/2){\cal R}_2\right]. 
\end{eqnarray}
%
Since ${}^{(4)}R_{\mu\nu} \simeq 3H^2 q_{\mu\nu}$, we have ${\cal R}_2\simeq\frac{3}{2}H^4$.
Thus we finally obtain the same expression as 
Eq. (\ref{diff}). 
From the above analysis, we 
conclude that the discrepancy between the holographic entropy $S_{\rm ent}$ and 
the de Sitter entropy $S_{\rm area}$ comes from the higher derivative correction to the 
field equations on the brane. 

\section{Higher curvature terms in the bulk}

Now let us discuss Ryu and Takayanagi's proposal in theory 
with higher derivative terms in the bulk. 
Although the low-energy effective action of string 
theory is the Einstein-Hilbert action, it will receive the corrections at high energies,
which will appear as a series 
of higher derivative terms. For simplicity,  
we concentrate on the Gauss-Bonnet correction term. 

There would be no reason why the current proposal of 
holographic entanglement entropy holds in this setup, 
because it is based on the original AdS/CFT duality without 
higher derivative corrections. Therefore, 
it is worth discussing the effect of higher derivative 
terms into the holographic 
entanglement entropy in the braneworld. 
See Ref. \cite{GBH} for related issues from different 
point of view. 

The action with the Gauss-Bonnet term is given by \cite{GB}
%
\begin{eqnarray}
S & = & \frac{1}{16\pi G_5}\int d^5x {\sqrt {-g}}
\Biggl[R-2\Lambda+\frac{\beta l^2}{4}(R^2-4R_{AB}R^{AB} \nonumber \\
& & +R_{ABCD}R^{ABCD}) \Biggr]
+\int d^4 x {\sqrt {-q}}(-\sigma + Q ),
\end{eqnarray}
%
where 
%
\begin{equation}
Q=2K+\beta l^2(J-2{}^{(4)}G^\mu_\nu K^\nu_\mu),
\end{equation}
%
with
%
\begin{equation}
J = -\frac{2}{3}K^\mu_\alpha K^\alpha_\beta K^\beta_\mu 
+K K^\alpha_\beta K^\beta_\alpha -\frac{1}{3}K^3. 
\end{equation}
%
Here $R_{ABCD}$, ${}^{(4)}G^\mu_\nu$ and $K^\mu_\nu$ are 
the five-dimensional Riemann tensor, the four-dimensional 
Einstein tensor and the extrinsic curvature on the brane, respectively. 
If the bulk is described by AdS spacetime with the curvature radius $l$, the 
bulk gravitational equations imply the following relation:
%
\begin{eqnarray}
\Lambda = -\frac{6}{l^2}+\frac{3\beta}{l^2}. 
\end{eqnarray}
%
The junction condition gives us the relation between $l$ and the brane tension $\sigma$ as 
%
\begin{eqnarray}
\frac{1}{l} {\rm coth}(r_0/l)\left[1-\frac{\beta}{3}+\frac{2\beta}{3{\rm sinh}^2(r_0/l)} \right]
=\frac{4\pi G_5}{3}\sigma. 
\end{eqnarray}
%
Using the above equations, the de Sitter entropy on the brane 
in this theory can be computed from the 
Euclidean path integral \cite{AM} as
%
\begin{eqnarray}
S_{{\rm E}}  &=& -I_{\rm E} \nonumber\\
&=& \frac{\pi l^3}{G_5}
[ 
(1+\beta) {\rm sinh}(r_0/l) {\rm cosh}(r_0/l)
\nonumber\\&&\qquad
-(1-3\beta) r_0/l]. 
\end{eqnarray}
%
In the limit of $r_0/l \gg1$, this reduces to
%
\begin{equation}
S_{{\rm E}} \simeq (1+\beta) \frac{\pi l}{G_5 H^2}.
\end{equation}
%

If the entanglement entropy for the Gauss-Bonnet theory
is given by the same formula proposed by Ryu and Takayanagi, we have the same 
result in the previous subsection because the bulk metric has the identical form:
%
\begin{equation}
S_{\rm ent}= \frac{\Omega_2 l^2}{2G_5}\int^{r_0}_0 dr~{\rm sinh}^2(r/l) 
\simeq \frac{\pi l}{G_5 H^2}. 
\end{equation}
%
Thus there is a discrepancy by the factor $(1+\beta)$ between 
the two entropies\footnote
{The relation between $G_5$ and $G_4$ is not directly related to 
our current consideration. It is known that it has the complicated 
scale dependence. See Ref. \cite{MS} for the detail of gravity on 
a de Sitter brane.}. This result suggests 
that the holographic entropy proposal based on the minimal surface is not 
applicable to higher derivative theories. 
Although we have observed the discrepancy here,
we do not intend to draw a negative conclusion
for the geometrical description of the entanglement 
entropy. 
In particular, because both $S_{\rm E}$ and $S_{\rm ent}$
have the same dependence on $H$ and $l$, the result
might indicate that the simple modification of the definition $S_{\rm ent}$
could give the exact value of the de Sitter entropy, although
further studies are needed to obtain the definite conclusion. 

\section{Summary and discussion}

In this paper we have studied the holographic entanglement entropy 
for a de Sitter braneworld embedded in an AdS bulk spacetime. 
We showed that 
the holographic entanglement entropy $S_{\rm ent}$ 
exactly agrees with the de Sitter entropy on the brane $S_{\rm E}$
evaluated by the Euclidean path integral. On the other hand, $S_{\rm ent}$ 
is not equal to the entropy $S_{\rm area}$ 
calculated from the horizon area formula on the brane in general 
because the effective theory on the 
brane deviates from the conventional Einstein theory. 
However, they coincide with each other in 
the limit of $r_0/l \gg 1$. 
This result is reasonable because in the limit the brane is 
sent off to the conformal boundary of AdS and 
then the brane theory decouples from bulk gravity, reproducing Einstein gravity on the brane.

We have also discussed the coverage of the holographic entanglement entropy 
for the braneworld model  
with the Gauss-Bonnet term. 
The current formulation was found to be not applicable for this corrections, 
and so 
the reformulation of the holographic entanglement entropy is awaited.

In this paper, we have concentrated on the de Sitter braneworld for simplicity. 
We expect, however, that the present discussion can be generalized to 
the Friedmann-Robertson-Walker (FRW) braneworld. 
We can compute the holographic entropy once the minimal surface is given, but the problem is 
that the entanglement entropy on the FRW brane is obscure due to the dynamical 
nature of the FRW spacetime.


\acknowledgements

We are grateful to R. Emparan for useful comments and suggestions. 
The work of TS was supported by Grant-in-Aid for Scientific 
Research from Ministry of Education, Science, Sports and Culture of 
Japan(No.13135208, No.14102004, No. 17740136 and No. 17340075), 
the Japan-U.K. and Japan-France Research  Cooperative Program.
TK is supported by the JSPS(No.~01642).
The work of HY was partially supported by a Grant for The 21st Century
COE Program (Holistic Research and Education Center for Physics 
Self-Organization Systems) at Waseda University.

\appendix

\section{Minimal surfaces}

In this section, we will show that $\rho=H^{-1}$ is the minimal surface in the 
bulk as well as on the brane. 
We begin with the review of the well-known result that 
the horizon is the minimal surface on static slices. Hereafter, we simply 
call the surface with zero expansion rate the horizon. 

The future directed null vector $n_\pm^a$ of out/in going null geodesics are 
locally decomposed as $n^a_\pm=t^a \pm s^a$, where $t^a$ is 
the future directed timelike vector and $s^a$ is the 
unit normal spacelike vector. Then the horizon is located at the zero 
expansion rate of null geodesic congruences $\theta_\pm$ in static spacetimes. 
The expansion rate is written as 
%
\begin{eqnarray}
\theta_\pm = {}^{(n)}K-{}^{(n)}K_{ab}s^a s^b \pm {}^{(n-1)}k, 
\end{eqnarray}
%
where ${}^{(n)}K_{ab}=(g_a^{~c}+t_a t^c)\nabla_c t_b =h_a^c \nabla_c t_b$ and 
$k=h^{ab}\nabla_a s_b$. Let $t^a$ be a normal vector to the static slices. With 
$K_{ab}=0$, we obtain  
%
\begin{eqnarray}
\theta_\pm = \pm {}^{(n-1)}k. 
\end{eqnarray}
%
Therefore, $k=0$ if $\theta_\pm=0$. 

The induced metric of $T={\rm constant}$ surface is 
%
\begin{multline}
h_{ab}dx^a dx^b  =  dr^2+ (l H)^2\sinh^2(r/l)\times 
\\
\left[(1-H^2\rho^2)^{-1}d\rho^2 
 +\rho^2d \Omega_{n-2}^2\right].
\end{multline}
%
On the brane, the minimal surface is the cosmological horizon at $\rho=H^{-1}$. 
Our task is to find the {\it bulk} minimal surface with the boundary at the 
cosmological horizon on the brane. 

Let $\rho=h(r)$ be the orbit of the bulk minimal 
surface, the unit normal vector to the surface is 
given by 
%
\begin{multline}
s^a =  \frac{1}{\sqrt{h_{,r}^2+\frac{1-H^2\rho^2}{(l H)^2\sinh^2(r/l)}}}\times
\\
\left[-h_{,r} (\partial_r)^a 
+ \frac{1-H^2\rho^2}{(l H)^2\sinh^2(r/l)}(\partial_\rho)^a \right].
\end{multline}
%
From the definition of the minimal surface, the trace of the extrinsic curvature vanishes:
%
\begin{eqnarray}
{}^{(n-1)}k & = & {}^{(n)}D_a s^a \nonumber \\
 & = & \partial_r s^r+\partial_\rho s^\rho+
\left(\Gamma^r_{rr} +\Gamma^\rho_{\rho r} +\Gamma^A_{A r} \right)s^r \nonumber \\
& & + \left(\Gamma^r_{r\rho} +\Gamma^\rho_{\rho\rho}
+\Gamma^A_{A\rho} \right)s^\rho =0, \label{minimal}
\end{eqnarray}
%
where ${}^{(n)}D_a$ is the covariant derivative with respect to $h_{ab}$ and $A,B$ denote 
the coordinate of $(n-2)$-dimensional unit sphere. 

We first show $s^r|_{\rm brane} \propto h'|_{\rm brane}=0$ 
which indicates that the bulk minimal surface is orthogonal to 
the brane. To see this, we take the integration around 
the neighborhood of the brane along the 
coordinate of the extra dimension $r$. Then we obtain 
%
\begin{eqnarray}
\lim_{r \to r_0-0} s^r = \lim_{r \to r_0+0} s^r. \label{even}
\end{eqnarray}
%
Due to the $Z_2$ symmetry, we see
%
\begin{eqnarray}
\lim_{r \to r_0-0}h' =- \lim_{r \to r_0+0}h' 
\end{eqnarray}
%
and this implies 
%
\begin{eqnarray}
\lim_{r \to r_0-0} s^r =- \lim_{r \to r_0+0} s^r. \label{odd}
\end{eqnarray}
%
Eqs. (\ref{even}) and (\ref{odd}) yield 
%
\begin{eqnarray}
\lim_{r \to r_0-0} s^r = \lim_{r \to r_0+0} s^r =0. 
\end{eqnarray}
%
Usually the bulk minimal surface with the fixed boundary on the brane is not 
orthogonal to the brane. However, the $Z_2$ symmetry requires the orthogonality. 
It is reminiscent of the fact  
that the bulk minimal surface with the free boundary bounded in 
the brane is orthogonal to the brane. 

Let us take the $(n-1)$-dimensional surface of $h(r)={\rm constant}$. Then $s^r 
\propto h_{,r} =0$ on the surface. 
Eq. (\ref{minimal}) and the normal vector $s^a$ become 
%
\begin{equation}
{}^{(n-1)}k= \partial_\rho s^\rho+\left(\Gamma^\rho_{\rho\rho}
+\Gamma^A_{A\rho}
\right)s^\rho \label{minimal2}
\end{equation}
%
and
%
\begin{equation}
s^a=\sqrt{\frac{1-H^2\rho^2}{(l H)^2\sinh^2(r/l)}}(\partial_\rho)^a. \label{normal}
\end{equation}
%
Using Eq. (\ref{normal}), Eq. (\ref{minimal2}) can be rewritten as
%
\begin{equation}
{}^{(n-1)}k= \Gamma^A_{A \rho}s^\rho =\frac{n-2}{\rho}s^\rho.
\end{equation}
%
Therefore, from the expression of $s^\rho$, the $\rho=H^{-1}$ surface in the bulk is turned out to be 
the bulk minimal surface.

\end{document}